\documentclass[
reprint,
superscriptaddress,
preprintnumbers,
 amsmath,amssymb,
pra
floatfix]{revtex4-1}

\usepackage{braket}
\usepackage{comment}
\usepackage{graphicx}
\usepackage{float}
\usepackage{tikz}
\usepackage{dcolumn}
\usepackage{bm}
\usepackage{hyperref}

\usepackage{mathtools}
\usepackage [english]{babel}
\usepackage [autostyle, english = american]{csquotes}
\usepackage{comment}
\MakeOuterQuote{"}
\DeclareMathOperator\arctanh{arctanh}
\begin{document}

\title{Boson Sampling with Gaussian input states: toward efficient scaling and certification}

\author{Raphael A. Abrahao}
\email{rakelabra@bnl.gov}
\affiliation{
Brookhaven National Laboratory, Upton, NY 11973, USA
}
\affiliation{
Centre for Engineered Quantum Systems, School of Mathematics and Physics, University of Queensland, St Lucia, Brisbane, Queensland 4072, Australia
}

\author{Arman Mansouri}
\affiliation{
Department of Physics, University of Ottawa,
25 Templeton Street, Ottawa, Ontario, Canada K1N 6N5 
}

\author{Austin P. Lund}
\email{austin.lund@gmail.com}
\affiliation{
Centre for Quantum Computation and Communication Technology, School of Mathematics and Physics,
University of Queensland, St Lucia, Brisbane, Queensland 4072, Australia
}
\affiliation{Dahlem Center for Complex Quantum Systems, Freie Universit\"at Berlin, 14195 Berlin, Germany}

\date{\today}

\begin{abstract}
A universal quantum computer of large scale is not available yet, however, intermediate models of quantum computation would still permit demonstrations of a quantum computational advantage over classical computing and could challenge the Extended Church-Turing Thesis. One of these models based on single photons interacting via linear optics is called Boson Sampling. Although Boson Sampling was demonstrated and the threshold to claim quantum computational advantage was achieved, the question of how to scale up Boson Sampling experiments remains. To make progress with this problem, here we present a practically achievable pathway to scale Boson Sampling experiments by combining continuous-variable quantum information and temporal encoding.  We propose the combination of switchable dual-homodyne and single-photon detections, the temporal loop technique, and scattershot-based Boson Sampling. We detail the required assumptions for concluding computational hardness for this configuration.  Furthermore, this particular combination of techniques moves towards an efficient scaling and certification of Boson Sampling, all in a single experimental setup.
\end{abstract}

\maketitle


\section{\label{sec:level1}Introduction}
Boson Sampling is a model of intermediate---as opposed to universal---quantum computation, initially proposed to confront the limits of classical computation compared to quantum computation~\cite{AA}. An efficient classical computation of the Boson Sampling protocol would support the Extended Church-Turing Thesis "which asserts that classical computers can simulate any physical process with polynomial overhead"\cite{harrow17}, i.e., polynomial time and memory requirements.  But an efficient classical algorithm for Boson Sampling would also imply that the Polynomial Hierarchy  of complexity classes, which is believed to have an infinite number of distinct levels, would reduce (or ``collapse'') to just three levels. Consequently, a computer scientist could not simultaneously support the Extended Church-Turing Thesis and an infinite structure of the Polynomial Hierarchy.  Hence, one is cornered into a position that either a fundamental change in our understanding of computational complexity is needed or quantum-enabled algorithms must be able to perform some tasks efficiently that cannot be performed efficiently on a classical computer. 

Even in the absence of a full-scale quantum computer, a physically constructed Boson Sampling device could outperform a classical device sampling from the same distribution \cite{harrow17,lund2017,review_road_QCS,Brod2019_review,Preskill2018_NISQ}. The first experimental demonstration of quantum computational advantage was published in October 2019 for a sampling problem using 53 qubits in a superconducting circuit architecture \cite{google_q_supremacy}, while additional analysis of the performance can be found in Ref.~\cite{LeveragingArxiv}. Recently, quantum computational advantage was demonstrated by two other groups: one using photons with 50 single-mode squeezed states in a 100-mode linear optical network \cite{q_advantage_2020_jwpan}, and the other one using 216 modes and temporal encoding in a 3D topology \cite{Madsen2022}. We note, however, a discussion whether these constitute a useful quantum computational advantage. This critique points out a need of a quantum computational advantage beyond the scope of computational complexity, while others favor the view that testing computational complexity can constitute a meritorious goal in its own right.

\begin{figure}
\includegraphics[width=1.0 \columnwidth]{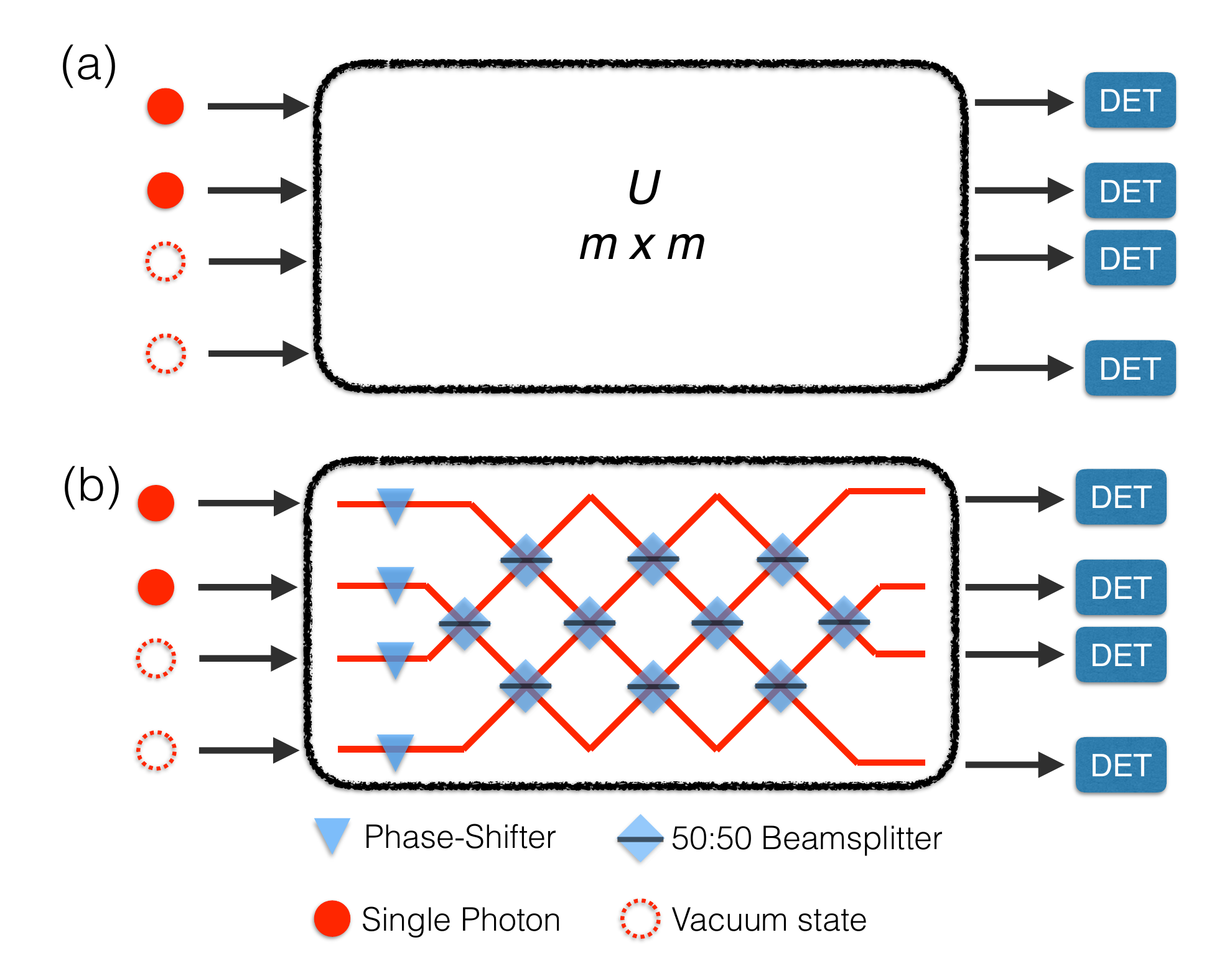}
\caption{(a) Conceptual schematics for the Boson Sampling protocol: $n$ photons, e.g. $n=2$, enter a linear optical network whose action is represented by a $m$ x $m$ unitary matrix ($U$), e.g. $m=4$, which relates the input amplitudes to the output amplitudes. The $m-n$ remaining inputs are considered to be in a vacuum state. The outputs are recorded using single-photon detectors (DET). (b) Conceptual implementation of Boson Sampling. Here the unitary $U$ is implemented in spatial modes using phase-shifters and 50:50 beamsplitters.}
\label{BS_general}
\end{figure}

In a simplified view, an implementation of the Boson Sampling protocol can be summarized as follows: $n$ indistinguishable single photons are inputs into the ports of an $m$ mode linear optical network, represented by a unitary matrix ($U$), and single-photon detection is performed at the output ports (Fig. \ref{BS_general}). Any alleged Boson Sampling device must give samples from this output distribution for any given $U$. The computational hardness is expressed as the scaling of runtime in terms of the number of input photons $n$.  The classical hardness has been proven when $m$ scales as $n^3$ in the general case. It has also been proven when $m$ scales as $n^2$ and the linear operation can be represented by an orthogonal matrix \cite{Chabaud2021Quantum}.

Implementations of Boson Sampling have been successfully demonstrated, initially using single-photon pairs from Spontaneous Parametric Down Conversion (SPDC) and later using single photons from Quantum Dots \cite{BroomeBS,SpringOX-BS,tillmannVienna,CrespiRome,ScattershotRome, LoredoBS,JWP_BS_2017, JWPan_BS_2018_PRL,flamini2018photonic}. 
A more advanced implementation achieved the threshold of quantum computational advantage using 50 single-mode squeezed states as inputs for an optical network with 100 modes \cite{q_advantage_2020_jwpan}. Another important milestone was the photonic implementation of programmable Boson Sampling with 216 squeezed modes entangled with three-dimensional connectivity using a time-multiplexed network and photon-number-resolving detection \cite{Madsen2022}. This work claimed a runtime advantage over 50 million times as extreme as reported by earlier photonic machines.

The question is then how to keep scaling up Boson Sampling experiments. Three factors are currently contributing against quantum demonstrations of Boson Sampling: (a) better classical algorithms which move the threshold of quantum computational advantage to greater number of input single photons \cite{nevilleBristol2017,Clifford&Clifford, Clifford&Clifford_2020}, e.g. the classical algorithm of Neville \textit{et al.} \cite{nevilleBristol2017} solved the Boson Sampling problem with 30 photons in a standard computer efficiently; (b) difficulties on the scaling of the preparation of multiple single photons (i.e $\ket{1}^{\otimes n}$) \cite{kaneda2019,loredo2016scalable,lenzini2017active,uppu2020scalable}; and (c) scaling of photon losses in the linear optical network \cite{Patron_Renema,Oszmaniec_Brod_err, Renema_classical_simulability_arxiv}.  Different theoretical directions have been pursued to study different types of quantum states~\cite{Lund2014,gaussian_BS_prl_2017,Chabaud2017} and measurements~\cite{Lund2017measurement,Chakhmakhchyan2017} to circumvent these problems. However, the challenge of scaling preparation and certification remains a topic of interest.

\begin{figure*}
\includegraphics[width=1.5 \columnwidth]{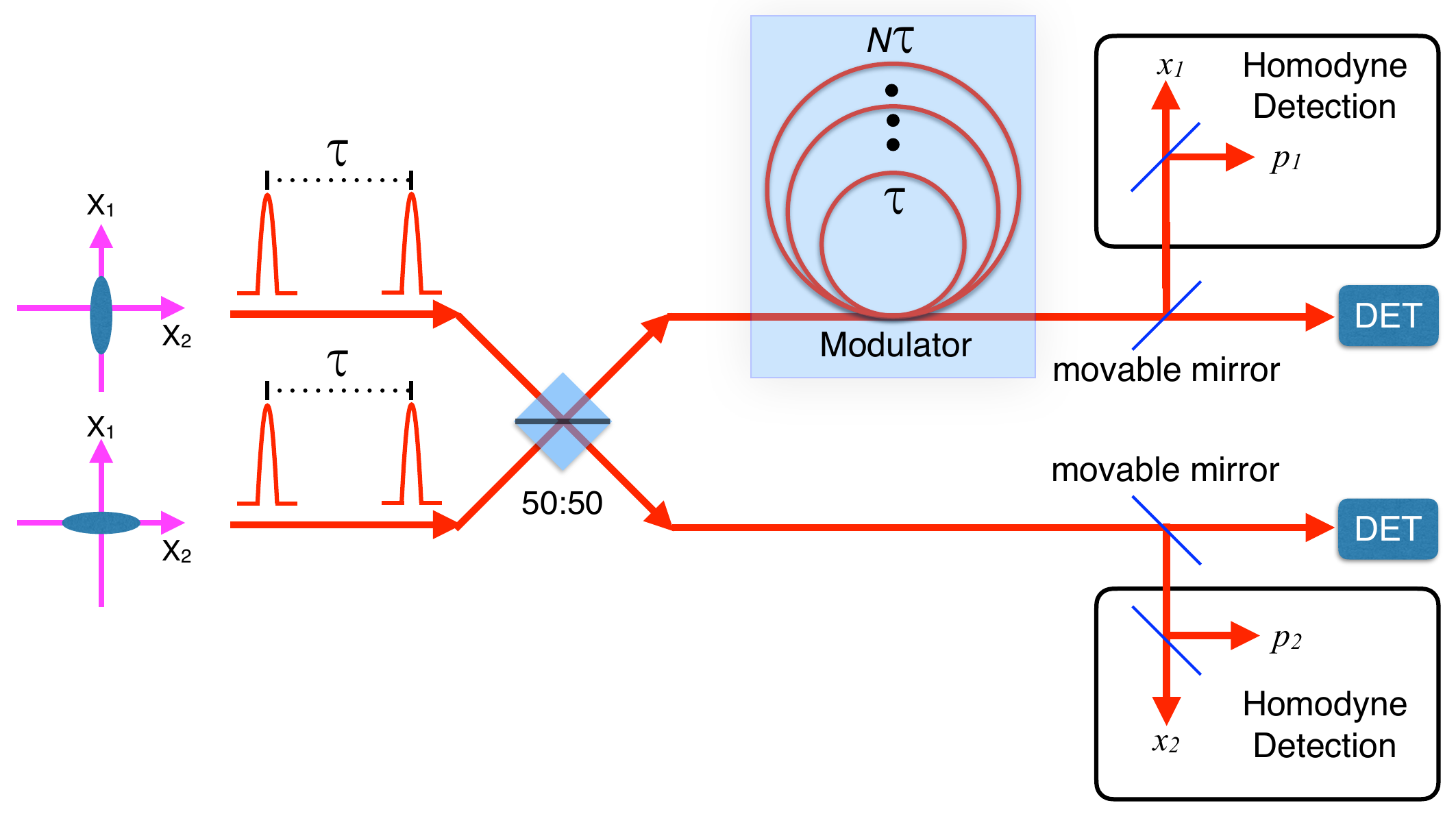}
\caption{(a) Schematics for Continuous-Variable Boson Sampling using temporal encoding. Input states are pulsed-Gaussian-squeezed states in an orthogonal basis, whose temporal difference is given by $\tau$. The 50:50 beamsplitter interferes these two inputs. The photons in the upper arm enter a "Modulator", while photons in the bottom arm propagate freely. In each arm, movable mirrors are placed to direct light in a characterization stage using Homodyne Detection. After the characterization, the movable mirrors must be removed and let light go to the corresponding single-photon detectors (DET) to generate the output (samples) for the Boson Sampling protocol, with this last stage, i.e. the projection of Gaussian states onto Fock basis, done similarly to the scattershot case \cite{Lund2014}. The "Modulator" can be implemented for example as in Ref.~\cite{Madsen2022}.}
\label{CV_BS_v2}
\end{figure*}

Implicit in the context of the implementation of Boson Sampling devices is that the operation of the device does indeed satisfy the theoretical requirements for an advantage over classical computational devices.  The condition presented in Ref.~\cite{AA} was on the total variation distance for the output distribution compared with the ideal distribution.  The standard view is aligned with that of cascaded quantum networks (e.g. constructions like those of Ref.~\cite{Gardiner1993,Carmichael1993}) where the performance of individual components are characterised and their performance as a network is extrapolated.  This approach, though conceptually straightforward, suffers from exponential sensitivity on the performance of interconnects between components.  Ideally, the performance of a Boson Sampling device could be characterised in situ.  A number of efficient schemes have been proposed to perform these kinds of validation tests, such as tests against uniform distributions~\cite{Aaronson2014}, particle distinguishability~\cite{Shchesnovich2016}, mean-field approximations~\cite{Tichy2014}, and others (see Ref.~\cite{Brod2019_review} for a detailed list of such tests). The drawback of most of the schemes to date is that they test a particular property of the output distributions, rather than verifying that they confirm the output distributions meet the theoretical requirements for a quantum advantage with Boson Sampling. Also, an efficient verification method for Boson Sampling with single-photon input states has recently been proposed \cite{Chabaud2021Quantum} \footnote{The paper \cite{Chabaud2021Quantum} cites a previous version of the current paper previously available on arXiv.}.  

Due to the exponential sized sampling space at the output, it is a difficult problem in general to find many individual conditions required to be simultaneously satisfied by a non-ideal output distribution. For the exponentially sized sample space, an exponentially sized ensemble is required to directly estimate the total variation distance without any prior assumptions (see Ref.~\cite{Hangleiter2019} for properties of distributions needed to avoid this issue). 

At this point in time, Boson Sampling faces significant challenges in both building and characterising scaled devices. Motivated by these constraints, we propose a method to scale Boson Sampling experiments using continuous-variable quantum states and time-bin encoding~\cite{weedbrook2012gaussian,Furusawa_10000,Furusawa_one_million}. Our analysis of this proposal accounts for finite squeezing and given some reasonable assumptions hold, operational performance can be characterized efficiently.  Our scheme presents a pathway toward an efficient way to implement Boson Sampling experiments, and also toward an efficient way to certify it using the same experimental setup. Here, by changing the detection type and making the assumption that the state before detection is Gaussian, we certify the total variation distance via the state covariance matrix. This is a quantity that has a polynomial dependence on the input size, and hence this statistical distance can be efficiently estimated.

It is noteworthy that the current search for applications of Boson Sampling goes beyond the scope of computational complexity. For instance, Boson Sampling has been adapted to simulate molecular vibrational spectra \cite{molecularBS_theory,molecularBS_exp} and may be used as a tool for quantum simulation \cite{aspuru2012photonic,georgescu2014quantum}.  Other Boson Sampling-inspired applications are the verification of NP-complete problems \cite{verifying_NP}, quantum metrology sensitivity improvements~\cite{MORDOR}, and a quantum cryptography protocol \cite{BS_quantum_crypt}.

\section{Proposal}

Here we propose a technique to scale Boson Sampling experiments based on continuous variables (CV) and temporal encoding.  In the CV case, the information is encoded in the quantum modes of light, specifically in the eigenstates of operators with a continuous spectrum \cite{kok2010introduction}. Continuous-variables quantum information has achieved impressive results. An initial report of 10,000 entangled modes in a continuous-variable cluster mode \cite{Furusawa_10000} was later upgraded to one million modes \cite{Furusawa_one_million}.  Some of these systems were conceived to perform measurement-based quantum computation, and here we show they can be adapted to Boson Sampling. Moreover, while some of the theoretical work for measurement-based quantum computation assume unrealistically infinite squeezing, we only require finite squeezing here. The world record for detected squeezed light is 15 db \cite{Roman_15db}, while 20.5 db of squeezing would give rise to conditions for fault-tolerant quantum computation under the Gottesman-Kitaev-Preskill encoding \cite{Menicicci2014,GKP_encoding}. 

The work of Lund \textit{et al.} \cite{Lund2014} (a.k.a. Scattershot Boson Sampling) demonstrated that Gaussian states can be used as inputs in Boson Sampling experiments and only bounded squeezing is necessary, provided each output is projected onto the number basis by single-photon detection. It is important at this point to emphasise that the specific task of Scattershot Boson Sampling requires the generation of a different distribution to standard Boson Sampling. One result presented in Ref.~\cite{Lund2014} was to show that sampling from this alternative distribution is also a hard problem when employing classical computing resources. The single photon Boson Sampling distribution is contained within the Scattershot Boson Sampling distribution, but this is merely used as part of the proof for computational hardness. The requisite task is the efficient generation of samples from a Gaussian state measured in the Fock basis without any further processing. 

A detailed discussion of how much squeezing is necessary for Scattershot Boson Sampling experiments can be found on Ref.~\cite{Lund2014}. Interestingly, the authors \cite{Lund2014} showed that for a two-mode squeezer, like SPDC~\cite{SPDC_general,Farella2023,PRL_Oxford_spdc2008}, there is a trade-off between the strength of the SPDC (linked to $\chi$) and the most likely number of photons detected, represented by the variable $n$.  This indicates that Scattershot Boson Sampling experiments that are performed with fewer photons would require higher $\chi$ levels: 
\begin{equation}
P(n)=\underbrace{\binom{m}{n} \underbrace{ \chi^{2n}(1-\chi^2)^{m}}_{\text{standard Boson Sampling}}}_{\text{scattershot Boson Sampling}},
\label{Eq_scattershot}
\end{equation}
where $P(n)$ is the probability of detecting $n$ photons and this probability is locally maximised when 
\begin{equation}
\chi = \sqrt{\frac{n}{m+n}}.
\end{equation}
This maximum probability is lower bounded by $1/\sqrt{n}$ if $m\geq n^2$ and $n>1$.  In this regime, $\chi$ decreases as $n$ increases and when taking $m=n^2$ at $n=8$ only $3$dB of squeezing is required to achieve this optimal probability. See \footnote{In the case of Scattershot Boson Sampling \cite{Lund2014}, $m$ refers to the number of two-mode squeezers, i.e., the number of SPDC single-photon sources. In that paper, the condition $m=n^2$ is imposed. In the present work, we use a more general definition: $m$ is simply the number of modes in the linear optical network represented by a $m$ x $m$ unitary matrix ($U$).} for clarification of the notation for Scattershot Boson Sampling.

More recently, a new variant of Boson Sampling was proposed in which the Boson Sampling protocol is formulated in terms of the Hafnian of a matrix, a more general function belonging to the \#P class, and was proven for the case of exact sampling. Moreover, it was shown in Ref.~\cite{Grier2022} that under the standard Anti-Concentration and Permanent-of-Gaussian conjectures, there is no classical algorithm to efficiently sample from the ideal Hafnian-based distributions, even approximately, unless the polynomial hierarchy collapses. This protocol is called Gaussian Boson Sampling \cite{gaussian_BS_prl_2017, gaussian_BS_arxiv_supp,deshpande2022quantum}.  In this present work, we use the terminology \textit{scattershot Boson Sampling} and \textit{Gaussian input Boson Sampling} as synonyms. A demonstration of scattershot, Gaussian, and standard Boson Sampling using integrated optics and single photons from Spontaneous Four Wave Mixing sources can be found in Reference \cite{BS_integrated_bristol_2019}.

\subsection{Scaling}

Now, we will analyse the scaling of our Boson Sampling experimental proposal. Consider two pulsed-squeezed-light sources, with $\tau$ time interval between subsequent pulses, where these two states interfere in a 50:50 beamsplitter, followed by a controllable delay, where a pulse can be delayed by $N\tau$ before being released, Fig.(\ref{CV_BS_v2}). This may be a loop architecture \cite{Time-Bin_BS_motes} or a quantum memory, provided it returns the given input state with enough high fidelity. Here we will use the loop architecture. The modulator should implement the desired unitary operation $U$ by interfering delayed pulses. At the end of each spatial path, two types of measurement can be performed.  Either the light is sent to single-photon detectors to record the samples (output) for Boson Sampling, or the light is directed towards a homodyne detection \cite{leonhardt1997measuring,Bachor&Ralph} setup used to characterize the output state, including the output state of the optical network.  

The use of time-bin (loop architecture) for Boson Samplings has been proposed and experimentally demonstrated \cite{Time-Bin_BS_motes, JWP_BS_2017}. This technique converts spatial modes into temporal modes. This gives a quadratic reduction in required spatial modes and provides a significant benefit in scaling to a large number of single photons as input states for the Boson Sampling protocol. However, the original time-bin proposal was conceived for the discrete case, i.e., manifold single photons. The certification in the discrete case, as addressed in the article \cite{Aolita2015} requires the tested state to be generated $\mathcal{O}(nm^n)$ times when verifying a process with $n$ photons in $m$ modes. So, for $m=n^2$, this is worse than an exponential growth. While the recent proposal in \cite{Chabaud2022} is an efficient verification scheme for Boson sampling with input single-photon states, it still relies on heterodyne measurements. Therefore, restricting both the input
and the measurement to discrete-variable does not constitute a
scalable approach for Boson Sampling when also considering
certification.

A significant benefit to our approach is that, under some reasonable assumptions, the operation of the sampling device can be characterized using the sampling state itself without the need for other probe input states. To achieve this, the following assumptions are needed: (i) the output state received by the single-photon detection is the same as that received by the homodyne detection, which is achievable by movable mirrors, for example as in the procedure given by \cite{webb2006homodyne}; (ii) the two squeezed input states are Gaussian and that the modulation network changes the states but leaves the output still in a Gaussian form, a standard Gaussian optics property; (iii) the output is fully characterized by a multi-mode covariance matrix, and (iv) the choice of when to make a sampling run and when to make a characterization run is irrelevant. In other words, the experimental setup is assumed stable, and the output will not change over the time as one changes between the two different measurement schemes. Finally, (v) all samples are generated under the independent and identically distributed (i.i.d.) assumption.

The first part of the schematics of Fig.(\ref{CV_BS_v2}), from the preparation of the two squeezed states to the 50:50 beamspliter to interfere these states, is a standard scheme in many CV quantum information protocol, as can be seen in References \cite{PhysRevApplied2021_Asavanant,larsen2021deterministic, larsen2019Science,asavanant2019Science}. The ``Modulator''can be implemented as a controllable temporal delay as utilized, for example, in Ref.~\cite{Madsen2022}. The novelty of our proposal is not the individual building blocks that compose it, but the proposal to use demonstrated building blocks to achieve a Boson Sampling platform with better scaling of preparation and better scaling of certification, as these two aspects will discussed. Moreover, the arguments for a better scaling of preparation of input state and better scaling of certification are also novel aspects.

A Gaussian output state can be fully characterized by the mean vector (which we will assume zero) and covariance matrix~\cite{kok2010introduction,Bachor&Ralph,weedbrook2012gaussian}.  For an $m$ mode state and $n$ detected photons, the number of possible photon number detection events scales as $(ne)^n$ (up to polynomial factors) in the regime $m=\mathcal{O}(n^2)$. This can be seen by considering the number of possible detection events $\binom{m+n+1}{n}$ and applying the asymptotic formula and Stirling's approximation. However, to describe a Gaussian state before the detection has occurred, only the number of entries in the covariance matrix for an $m$ mode state is required and this scales as $4 m^2$.  For the case of Gaussian input Boson Sampling (a.k.a. scattershot) where there are two groups of $m$ modes and $n$ photon detections, the size of the Fock basis detection sample space is $m^{2n}$, but the full covariance matrix for the state prior to detection will require $16 m^2$ entries.

Performing the characterization involves reconstructing the covariance matrix from the CV measurement samples.  The measurements chosen must be sufficient in number to estimate all elements of the covariance matrix, including terms involving the correlations between X and P in the same mode. To avoid repeated changes to measurement settings, we propose performing this by means of dual homodyne detection.  In a dual-homodyne arrangement, the signal mode is split at a 50:50 beamsplitter and both modes undergo a CV homodyne detection, one measured in X and the other in P.  This permits a simultaneous measurement of the X and P quadratures at the cost of adding $1/2$ a unit of vacuum noise to the diagonal elements of the state covariance matrix. So, if $\Sigma$ is the state covariance matrix, then the dual homodyne modes will see Gaussian statistics with a covariance matrix of $(\Sigma + I)/2$ (under units where the variance of vacuum noise is unity), where $I$ is the identity operator. Note that Ref.~\cite{Aolita2015} mentions the use of heterodyne detection, which gives the same statistics as dual homodyne detection. This covariance matrix can then be estimated by constructing matrix-valued samples from each sampling run. Let 
\begin{equation}
s_i = (x_{1,i},p_{1,i},x_{2,i},p_{2,i},\ldots,x_{m,i},p_{m,i})^T
\end{equation}
be a $2m$-dimensional real vector representing the $i^\mathrm{th}$ data sample from the dual-homodyne measurement, with the first subscript representing the mode to which the corresponding homodyne detector is attached. From this sample vector, a sample matrix can be formed from the outer product of the $s_i$
\begin{equation}
\xi_i = \frac{1}{C^2}s_i s_i^T,
\end{equation}
where $ C $ is a scaling factor. The reason for introducing this scaling factor will become apparent shortly. For now, one can think of it as being related to the maximum energy of the state. This sample matrix is a positive semi-definite matrix for all $i$.  The expectation value for each sample $\xi_i$ over the incoming Gaussian distribution is then
\begin{equation}
		\langle \xi_i \rangle = \frac{1}{2C^2}(\Sigma + I)
		\label{Eq_expectation_value} 
	\end{equation}
and so a sample average over $K$ samples
\begin{equation}
    \bar{\xi} = \frac{1}{K} \sum_{i=1}^K \xi_i
    \label{Eq_sample_average}
\end{equation}
will be an unbiased estimator for $(\Sigma + I)/(2C^2)$.

To see how close the sample average is to the true average, we apply the operator Chernoff bound \cite{wilde_2013, nielsen_chuang_book} (following the notation of Wilde \cite{wilde_2013},  Section 16.3). This provides a lower bound for the probability that the sample average does not deviate significantly from the expected value. Let $K$ be the number of sample matrices and $\bar{\xi}$ the sample average of $K$ samples as defined in Eq. \ref{Eq_sample_average}. The input state covariance matrix $\Sigma$ is positive definite, and we have
\begin{equation}
   \frac{1}{2C^2}(\Sigma + I) \geq \frac{1}{2C^2}I
\end{equation}
which is the expectation of each operator forming the sum in Eq. \ref{Eq_sample_average}.  For this situation, the operator Chernoff bound for any $0<\eta<1/2$ is given by
\begin{gather}
\text{Pr}\Big\{ \frac{(1-\eta)}{2C^2} (\Sigma + I) \leq \bar{\xi} \leq \frac{(1+\eta)}{2C^2} (\Sigma + I) \Big\} \nonumber \\
\geq 1 - 8 m e^{-K \eta^2 /(8C^2\ln 2)}.
\end{gather}
To then bound the probability for making a multiplicative estimate of $\Sigma$, the spectrum of $\Sigma$ needs to be bounded away from zero.

Let the parameter $b$ represent the variance of the quadrature for the maximum possible squeezing for the state being estimated.  This means that 
\begin{equation}
    \Sigma \geq bI.
    \label{Eq_spectum_bound}
\end{equation}
The Chernoff bound for the estimator $\bar{\xi}$ 
can be rewritten as
\begin{gather}
\text{Pr}\Big\{\frac{(1-\eta)\Sigma - \eta I}{C^2} \leq 2\bar{\xi} - \frac{I}{C^2} \leq \frac{(1+\eta)\Sigma + \eta I}{C^2}\Big\}  \nonumber\\
\geq 1 - 8 m e^{-K \eta^2 /(8C^2\ln 2)}.
\end{gather}
Then using the inequality in Eq. \ref{Eq_spectum_bound}, this can be written as
\begin{gather}
\text{Pr}\Big\{\frac{[1-\eta(1+b^{-1})]\Sigma}{C^2} \leq 2\bar{\xi} - \frac{I}{C^2} \leq \frac{[1+\eta(1+b^{-1})]\Sigma}{C^2}\Big\} \nonumber\\
\geq 1 - 8 m e^{-K \eta^2 /(8C^2\ln 2)}. \label{Chernoff_eq}
\end{gather}
This means the rewritten estimate $2 \bar{\xi} - I/C^2$ gives a multiplicative estimate of the scaled covariance matrix $\Sigma/C^2$. 

\begin{figure}[]
\centering
    \includegraphics[scale=0.58]{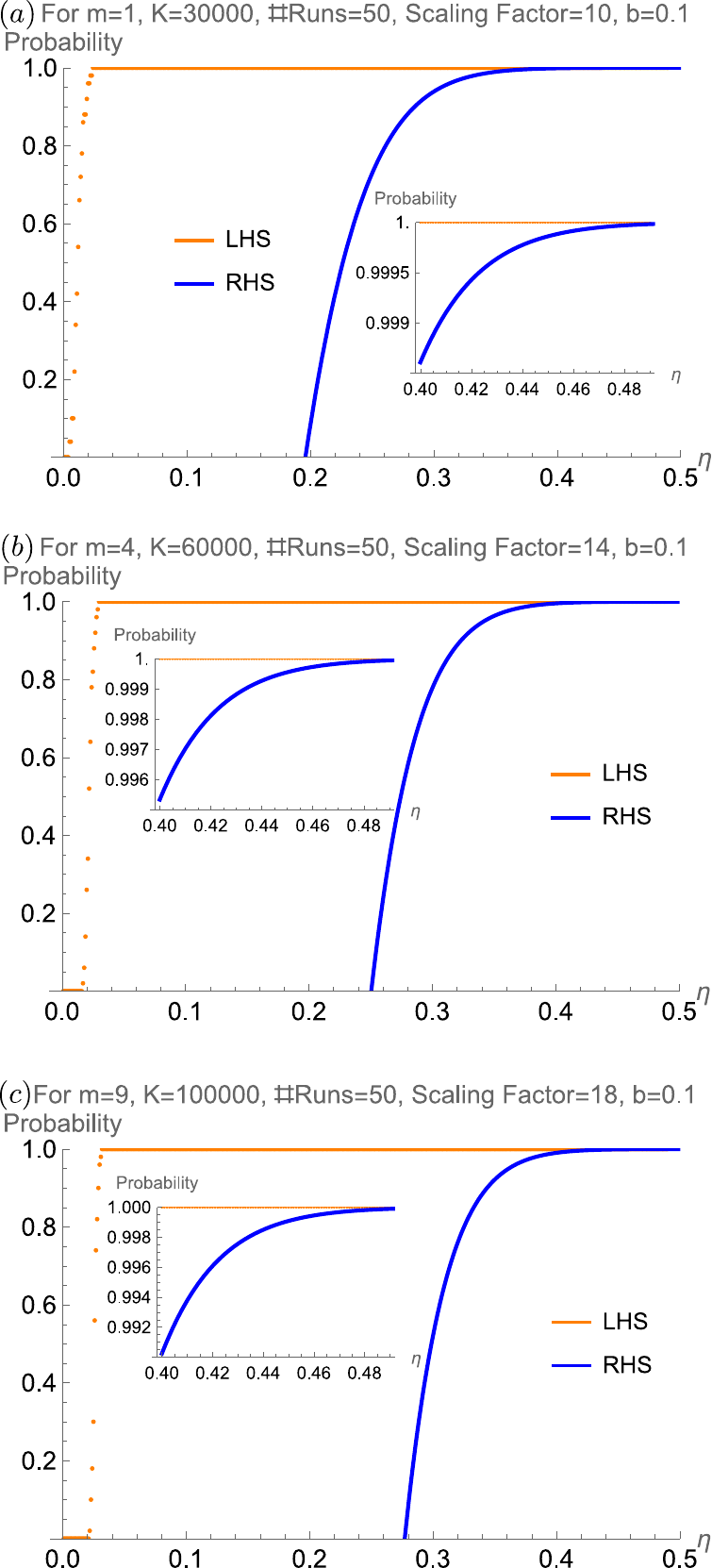}
    \caption{Plots of the probabilities on the left-hand side (LHS, shown as orange dots) and the right-hand side (RHS, shown as solid blue) of Eq. \ref{Chernoff_eq} as a function of $\eta$ (logarithmic scale) for the specified parameters. The covariance matrix $\Sigma$ is diagonal and has the specified diagonal entries. The number of runs refers to the number of trials used to calculate each probability. The insets show the region $\eta \geq 0.4$. The simulations are in agreement with Eq. \ref{Chernoff_eq}.}
    \label{Chernoff_bound_plots}
\end{figure}

The interpretation of Equation \ref{Chernoff_eq} is that the chance that the finite sample estimate of the covariance matrix deviates from the true value decays exponentially in the number of covariance matrix samples $K$ and the square of deviation permitted $\eta$, but depends linearly on $m$, the number of modes.  In our application of Gaussian input Boson Sampling, the value of $b$ is fixed, as too much squeezing can actually degrade performance.  So for fixed $\eta$, as the number of modes $m$ increases, the number of samples $K$ required to achieve the same probability bound in the operator Chernoff bound only grows logarithmically $\mathcal{O}(\ln m)$.

The validity of the operator Chernoff bound applied to the characterization of Gaussian states can be demonstrated numerically. We draw samples from a multivariate Gaussian distribution with mean-vector zero and covariance matrix $ (\Sigma + I)/2 $ (and scale the vectors by $C$). The vectors are used to construct the matrix samples $\xi_i$. The probability on the left-hand side of Eq. \ref{Chernoff_eq} is found by computing the sample average and finding how often the two conditions (in curly brackets) are satisfied. More details can be found in the Appendix.

Numerical simulations were performed for the cases of $m=1,4,9$ for several choices of $\Sigma$, $K$, and $b$. All the cases that were examined showed an agreement with Eq. \ref{Chernoff_eq}. In Fig. \ref{Chernoff_bound_plots}, we present some of the numerical results for the specified parameters. For $m=1$, Fig. \ref{Chernoff_bound_plots} is for the diagonal covariance matrix $\Sigma = \text{diag}(10, 0.1, 10, 0.1)$. The covariance matrices for higher $m$ can be found in the Appendix.

The figures presented here consider only the case of diagonal covariance matrices. These cases show the essential features of the bound as the eigenvalues of the covariance matrix contain the most critical information with respect to the Chernoff bound. Nevertheless, we present numerical simulations of non-diagonal covariance matrices, confirming this heuristic, in the Appendix. 

Note that a prerequisite for applying the operator Chernoff bound is that all samples $\xi_i$ have all of their eigenvalues in $[0,1]$ \cite{wilde_2013}. Although a smaller $C$ value (with other parameters being fixed) can provide a tighter bound, this prerequisite for applying the bound would be satisfied at an increasingly infrequent rate. For the purposes of these numerical simulations, the value of $C$ was chosen large enough such that this prerequisite is satisfied at a reasonable rate.
\subsection{Certification}

Finally, one would like to certify if the generated state is sufficient to perform the task at hand, that is Boson Sampling. For approximate Boson Sampling, one does not need to generate the state ideally but within some trace distance bound $\varepsilon$. Using the Fuchs-van de Graff inequality~\cite{fuchs1997cryptographic}, a trace distance is upper bounded by the fidelity by $\varepsilon\leq \sqrt{1-F}$.  A robust certification strategy is given by Aolita \textit{et al.} \cite{Aolita2015}, which tests if a fidelity lower bound (or equivalently maximum trace distance) holds between a pure Gaussian target state and a potentially mixed preparation state. Here, we argue that this result can also be employed for certification. In order to perform the certification, the Gaussian covariance matrix elements need to be estimated and manipulated with knowledge of the target pure state. This produces a bound of the fidelity, which can be used to test for appropriateness of the apparatus to perform Gaussian input Boson Sampling. The samples needed to achieve a fixed fidelity bound (or fixed trace distance) is higher than the Chernoff bound, and scales polynomially in $m$, the number of modes in the state being certified. Reference \cite{Aolita2015} indicates an $\mathcal{O}(m^4)$ scaling in the number of copies for the certification of Gaussian target states. However, calculating the coefficients for the scaling of this certification test for Boson Sampling with Gaussian input states reveals an $\mathcal{O}(m^5)$ scaling. In particular, for $m \geq 25$, the number of samples $K$ required for certification is estimated to be:
\begin{equation}
    K \geq (1.32158 \times 10^6) \ m^5 + (3.30395 \times 10^5) \ m^3 \label{eq:certification}
\end{equation}
The coefficients above can be obtained from the expressions presented in Ref.~\cite{Aolita2015} and by using properties of Gaussian states. Some assumptions regarding the target fidelity and success probability of the certification test were made to obtain the numerical coefficients. More details can be found in the Appendix. 

\begin{figure}[]
\centering 
\includegraphics[scale=0.6]{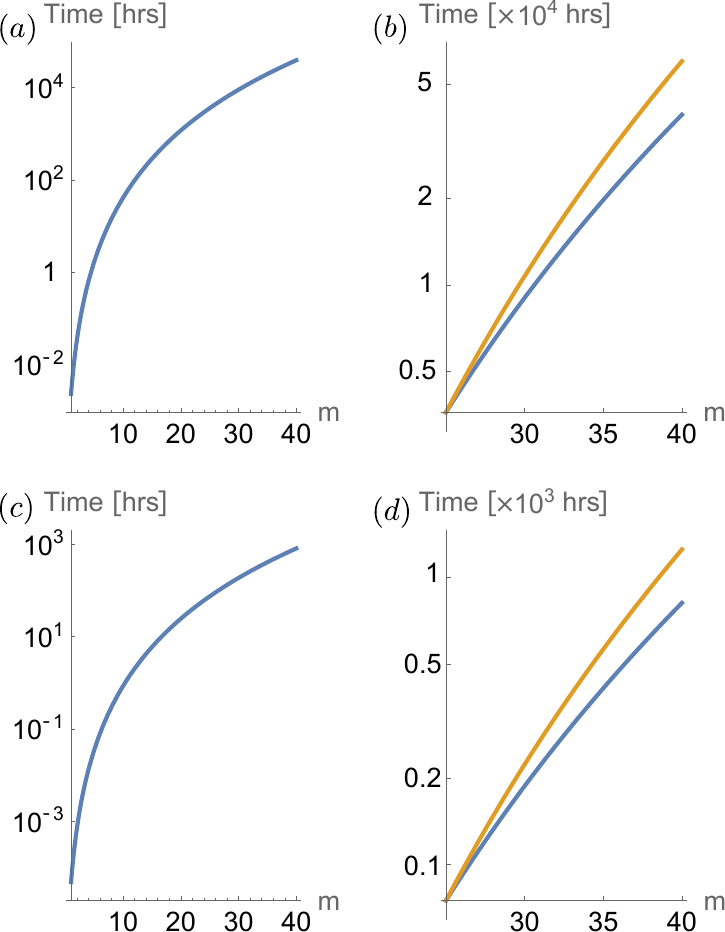}
\caption{Time estimates for the certification in \cite{Aolita2015} applied to Boson Sampling with Gaussian input states, shown as log-plots. (a) Estimated time when using the full constant (valid for all $m$) and assuming $25 \times 10^6$ measurements performed per second (Ref. \cite{asavanant2019Science}); (b) Estimated time from the simplified expression Eq.~\ref{eq:certification} (valid for $m\geq 25$) shown in orange, with the estimated time using the full constant shown in blue, when assuming $25 \times 10^6$ measurements per second; (c) Estimated time when using the full constant (valid for all $m$) and assuming $1.2 \times 10^9$ measurements performed per second (Ref. \cite{zhang2018IEEE}); (d) Estimated time from the simplified expression Eq.~\ref{eq:certification} (valid for $m\geq 25$) shown in orange, with the estimated time using the full constant shown in blue, assuming $1.2 \times 10^9$ measurements per second.}
\label{fig:certification}
\end{figure}

This certification process can require considerable amounts of data to scale, but the scaling with system size is polynomial, making the process more likely to be feasible. Figure \ref{fig:certification} (a) and (b) show the estimated time required for the certification in Ref.~\cite{Aolita2015} for a given number of input modes $m$, if one can perform $25 \times 10^6$ measurements per second, a number based on Ref.~\cite{asavanant2019Science}. This number indicates that the certification time for higher $m$ values remains prohibitive given established CV homodyne measurements. When instead considering a rate of $1.2 \times 10^9$ measurements per second, a value reported in Ref.~\cite{zhang2018IEEE} in the context of balanced homodyne detection for continuous-variable quantum information, one obtains better results for the estimated certification time, as expected. Figure 4 (c) and (d) show the estimated certification time based on the value of $1.2 \times 10^9$ measurements per second~\cite{zhang2018IEEE}.

Our proposed scheme is in contrast to the certification in the discrete variables model as addressed in the same article \cite{Aolita2015}, which requires $\mathcal{O}(nm^n)$ samples of the tested state to be generated when certifying a process with $n$ photons in $m$ modes. Our scheme, although not strictly scalable with the current quantum technology, suggests a better certification approach, and thus, a path toward scalable certification methods.

The fidelity bound achieved from this certification test is shown to be complete and robust to experimental imperfections. Completeness implies that if the target and prepared states are the same, the fidelity lower bound is in fact equal to the ideal fidelity between the prepared state $\rho_P$ and the target state $\rho_T$. Additionally, for a threshold fidelity $F_T$ , a fidelity buffer gap $\Delta<1-F_T$, and a maximal failure probability $\alpha>0$, with probability at least $1-\alpha$, this test both rejects every $\rho_P$ for which the $F(\rho_P, \rho_T) < F_T$ and accepts every $\rho_P$ for which $F(\rho_P, \rho_T)\geq F_T+\Delta$. The certification is robust against imperfections causing fidelity deviations as large as $1-(F_T+\Delta)$.

\subsection{Sampling}

After the stage of certification is finished, the movable mirrors must be removed to direct the light toward the single-photon detectors. Doing so, one is projecting the Gaussian states into the Fock basis, and thus obtaining the output of the Boson Sampling experiment, similarly as in the  Gaussian input Boson Sampling (a.k.a. scattershot) \cite{Lund2014}. Our proposed method greatly simplifies the numbers of required resources for scaling Boson Sampling experiments. Here we benefit from having well certified states, with certification growing polynomially as discussed above, and from having only two squeezed light sources, and thus simplifying the preparation of input states. Our method also requires fewer detectors. For instance, if one wishes to implement a 20 input single photons Boson Sampling, then it requires a $400 \times 400$ linear optical network, and therefore 400 single-photon detectors. Note that not obeying $m \gg n$, at least $m=n^2$, violates the mathematical assumptions upon which the approximate Boson Sampling problem is currently formulated. In our proposal, due to the time-bin implementation of the linear optical network, only 2 single-photon detectors are required. 

To elucidate some detection issues, two comments should be made. First, the comment about using only 2 single-photon detectors, made in the previous paragraph, is a very idealized case. In practice, it is likely that extremely fast detectors and/or some scheme of multiplexed detectors scheme would be used. Second, photo-number resolution is not required, although detectors like transition edge sensors can present such feature~\cite{Lita_NIST_OE_TES,TES_UQ_arxiv,NatPht_2022_TES,PRL2019_Lewis,gerrits_tes_intro}, or storage loop architecture~\cite{Ottawa_loop}, or limited number resolution in the case of nanowire detectors~\cite{nanowire_number_resolving,PhysRevApplied_number_resolving}.

In short, we exploit the properties of Gaussian states as long as we can, thus providing a better scheme for certification. The computational hardness come from dealing with the Fock basis, and thus, comes from projecting the final Gaussian states into the number (Fock) basis; this is achieved by the single-photon detectors (indicated as ``DET'' in Fig.~\ref{CV_BS_v2}). In other words, we use Gaussian properties when we need a simpler computational task (certification), while we use the Fock basis when we need a hard computational task (Boson Sampling samples).

\subsection{The role of imperfections}

An important consideration in the performance of any sampling device is the role that imperfections, originating from any process, have on the ability one has to make conclusions about the classical easiness or hardness of computing random samples. Crucially, the dominant imperfection using current technology is photon losses. The effects of photon losses are included within the approximate sampling requirements, i.e., how much one can deviate from the perfect sampling.  Unfortunately, this does not mean that losses can be neglected, as they will in many cases give rise to exponential scaling in the total variation distance between the lossless distribution and lossy distribution.  For example, a constant loss rate for each mode will induce an exponential scaling in the total variation distance as a function of the number of photons to be detected. It would seem that this implies that any level of loss would render classically hard Boson Sampling impossible, but this is not the case. The goal to demonstrate the quantum computational advantage only requires producing samples that are close to some distribution, given some $\varepsilon$ tolerance, that is hard for a classical computer to reproduce. Aaronson and Brod \cite{Aaronson_Brod_error} showed that the distribution generated from losing a fixed number of photons (i.e. not scaling with the number of photons) is hard for a classical computer to sample. A lossy Boson Sampling device will be close to this distribution at some scale. Unfortunately, the total variation distance for constant loss per mode will still asymptotically scale exponentially against this distribution with a fixed number of lost photons. Oszmaniec and Brod \cite{Oszmaniec_Brod_err} showed that if the total number of photons that remained after loss scaled as $\sqrt{n}$ of the number input photons $n$, then a simple distribution from sampling distinguishable bosons would satisfy the total variation distance requirement for approximate sampling. Hence an efficient classical computation could reproduce samples close to the required Boson sampling distribution, nullifying the quantum computational advantage. In a later article, Brod and Oszmaniec studied the case of nonuniform losses~\cite{brod_oszmaniec_nonuniform_losses}. 

Despite advances in understanding the effect of losses \cite{Arkhipov_error_analysis,rohde_ralph_error,Aaronson_Brod_error,L_Patron_error,rahimi_ralph_caves_sufficient,Oszmaniec_Brod_err, Patron_Renema, Renema_classical_simulability_arxiv,brod_oszmaniec_nonuniform_losses}, there remains a gap between the necessary and sufficient criteria for the hardness of approximate sampling with lossy Boson Sampling devices, which makes this an important open question. All implementations will be subject to imperfections, including the proposed implementation presented here. However, in this present work, we are primarily concerned with the experimental implications of scaling and certifying continuous-variables Boson Sampling. We expect that future results on the hardness of lossy Boson Sampling would be able to be incorporated into the proposal we present.

\section{Discussion}

Continuous-variables quantum information, particularly in the context of optical Gaussian states \cite{weedbrook2012gaussian}, has been put forward as an alternative for quantum computation. Due to the scaling aspects discussed in the previous sections, we point out that Boson Sampling can greatly benefit from the current optical CV technology \cite{Furusawa_10000,Furusawa_one_million, takeda2017universal, spie_travelling_wave_of_light, Roman_15db,furusawa_review_apl_2019}. In this sense, all the building blocks for this proposal have been successfully demonstrated and have good performance for scaling. The threshold for quantum computational advantage in the CV regime is currently uncertain and the subject of further investigation, but our Boson Sampling proposal certainly does not suffer from the scaling issues of discrete Boson Sampling, in particular the issue of preparing a large number of indistinguishable single photons. 

In summary, we revisited the motivation behind Boson Sampling and the experimental challenges currently faced. Photon losses and scaling of many input single photons are factors working against quantum implementations of Boson Sampling. These facts pose great challenges and a new scalable approach is desirable. Here we presented a new method to do so based on continuous variables and temporal encoding. Our method assumes finite squeezing and also provides a feasible way to perform the characterization of the input states and the certification of the Boson Sampling protocol, providing viable scaling as the system size increases. With this approach, we aim to propose a pathway to scale Boson Sampling experiments, and thus, make this intermediate mode of quantum computing more compelling for its foundational questions in Computer Science as well as its applications \cite{molecularBS_theory,molecularBS_exp,verifying_NP,MORDOR,BS_quantum_crypt}.

\section{\label{sec:level4}ACKNOWLEDGMENTS}
We thank Andrew G. White, T. C. Ralph, T. Weinhold, Nathan Walk, and Daniel Brod for helpful discussions. This work was supported by the U.S. Department of Energy QuantISED award, the Brookhaven National Laboratory LDRD grant 22-22. APL acknowledges support from BMBF (QPIC) and the Einstein Research Unit on Quantum Devices. This work was supported by the Australian Research Council Centre of Excellence for Engineered Quantum Systems (Grants No. CE170100009)
and the Australian Research Council Centre of Excellence for Quantum Computation and Communication Technology
(Grant No. CE170100012). This work was also supported by the CFREF award on  Transformative Quantum Technologies and the NSERC of Canada.
\\
\\
\textit{Disclosure}--APL consulted for Xanadu Quantum Technologies during this research and recently became an employee of Xanadu.

\bibliographystyle{ieeetr}

\bibliography{bib}

\begin{thebibliography}{10}

\bibitem{AA}
S.~Aaronson and A.~Arkhipov, ``The computational complexity of linear optics,'' {\em Theory of Computing}, vol.~9, no.~4, pp.~143--252, 2013.

\bibitem{harrow17}
A.~W. Harrow and A.~Montanaro, ``Quantum computational supremacy,'' {\em Nature}, vol.~549, no.~7671, p.~203, 2017.

\bibitem{lund2017}
A.~Lund, M.~J. Bremner, and T.~Ralph, ``Quantum sampling problems, bosonsampling and quantum supremacy,'' {\em npj Quantum Information}, vol.~3, no.~1, p.~15, 2017.

\bibitem{review_road_QCS}
C.~S. Calude and E.~Calude, ``The road to quantum computational supremacy,'' {\em arXiv:1712.01356}, 2017.

\bibitem{Brod2019_review}
D.~J. Brod, E.~F. Galvão, A.~Crespi, R.~Osellame, N.~Spagnolo, and F.~Sciarrino, ``{Photonic implementation of boson sampling: a review},'' {\em Advanced Photonics}, vol.~1, no.~3, pp.~1 -- 14, 2019.

\bibitem{Preskill2018_NISQ}
J.~Preskill, ``Quantum {C}omputing in the {NISQ} era and beyond,'' {\em {Quantum}}, vol.~2, p.~79, Aug. 2018.

\bibitem{google_q_supremacy}
F.~Arute, K.~Arya, R.~Babbush, D.~Bacon, J.~C. Bardin, R.~Barends, R.~Biswas, S.~Boixo, F.~G. Brandao, D.~A. Buell, {\em et~al.}, ``Quantum supremacy using a programmable superconducting processor,'' {\em Nature}, vol.~574, no.~7779, pp.~505--510, 2019.

\bibitem{LeveragingArxiv}
E.~Pednault, J.~A. Gunnels, G.~Nannicini, L.~Horesh, and R.~Wisnieff, ``Leveraging secondary storage to simulate deep 54-qubit sycamore circuits,'' {\em arXiv:1910.09534}, 2019.

\bibitem{q_advantage_2020_jwpan}
H.-S. Zhong, H.~Wang, Y.-H. Deng, M.-C. Chen, L.-C. Peng, Y.-H. Luo, J.~Qin, D.~Wu, X.~Ding, Y.~Hu, P.~Hu, X.-Y. Yang, W.-J. Zhang, H.~Li, Y.~Li, X.~Jiang, L.~Gan, G.~Yang, L.~You, Z.~Wang, L.~Li, N.-L. Liu, C.-Y. Lu, and J.-W. Pan, ``Quantum computational advantage using photons,'' {\em Science}, 2020.

\bibitem{Madsen2022}
L.~S. Madsen, F.~Laudenbach, M.~F. Askarani, F.~Rortais, T.~Vincent, J.~F.~F. Bulmer, F.~M. Miatto, L.~Neuhaus, L.~G. Helt, M.~J. Collins, A.~E. Lita, T.~Gerrits, S.~W. Nam, V.~D. Vaidya, M.~Menotti, I.~Dhand, Z.~Vernon, N.~Quesada, and J.~Lavoie, ``Quantum computational advantage with a programmable photonic processor,'' {\em Nature}, vol.~606, pp.~75--81, Jun 2022.

\bibitem{Chabaud2021Quantum}
U.~Chabaud, F.~Grosshans, E.~Kashefi, and D.~Markham, ``Efficient verification of {B}oson {S}ampling,'' {\em {Quantum}}, vol.~5, p.~578, Nov. 2021.

\bibitem{BroomeBS}
M.~A. Broome, A.~Fedrizzi, S.~Rahimi-Keshari, J.~Dove, S.~Aaronson, T.~C. Ralph, and A.~G. White, ``Photonic boson sampling in a tunable circuit,'' {\em Science}, vol.~339, no.~6121, pp.~794--798, 2013.

\bibitem{SpringOX-BS}
J.~B. Spring, B.~J. Metcalf, P.~C. Humphreys, W.~S. Kolthammer, X.-M. Jin, M.~Barbieri, A.~Datta, N.~Thomas-Peter, N.~K. Langford, D.~Kundys, J.~C. Gates, B.~J. Smith, P.~G.~R. Smith, and I.~A. Walmsley, ``Boson sampling on a photonic chip,'' {\em Science}, vol.~339, no.~6121, pp.~798--801, 2013.

\bibitem{tillmannVienna}
M.~Tillmann, B.~Daki{\'c}, R.~Heilmann, S.~Nolte, A.~Szameit, and P.~Walther, ``Experimental boson sampling,'' {\em Nature Photonics}, vol.~7, no.~7, p.~540, 2013.

\bibitem{CrespiRome}
A.~Crespi, N.~Spagnolo, P.~Mataloni, R.~Ramponi, E.~Maiorino, D.~Brod, R.~Osellame, F.~Sciarrino, C.~Vitelli, and E.~Galvao, ``Experimental boson sampling in arbitrary integrated photonic circuits,'' {\em Nature Photonics}, vol.~7, p.~545, 2012.

\bibitem{ScattershotRome}
M.~Bentivegna, N.~Spagnolo, C.~Vitelli, F.~Flamini, N.~Viggianiello, L.~Latmiral, P.~Mataloni, D.~J. Brod, E.~F. Galv{\~a}o, A.~Crespi, R.~Ramponi, R.~Osellame, and F.~Sciarrino, ``Experimental scattershot boson sampling,'' {\em Science Advances}, vol.~1, no.~3, 2015.

\bibitem{LoredoBS}
J.~C. Loredo, M.~A. Broome, P.~Hilaire, O.~Gazzano, I.~Sagnes, A.~Lemaitre, M.~P. Almeida, P.~Senellart, and A.~G. White, ``Boson sampling with single-photon fock states from a bright solid-state source,'' {\em Phys. Rev. Lett.}, vol.~118, p.~130503, Mar 2017.

\bibitem{JWP_BS_2017}
Y.~He, X.~Ding, Z.-E. Su, H.-L. Huang, J.~Qin, C.~Wang, S.~Unsleber, C.~Chen, H.~Wang, Y.-M. He, X.-L. Wang, W.-J. Zhang, S.-J. Chen, C.~Schneider, M.~Kamp, L.-X. You, Z.~Wang, S.~H\"ofling, C.-Y. Lu, and J.-W. Pan, ``Time-bin-encoded boson sampling with a single-photon device,'' {\em Phys. Rev. Lett.}, vol.~118, p.~190501, May 2017.

\bibitem{JWPan_BS_2018_PRL}
H.~Wang, W.~Li, X.~Jiang, Y.-M. He, Y.-H. Li, X.~Ding, M.-C. Chen, J.~Qin, C.-Z. Peng, C.~Schneider, M.~Kamp, W.-J. Zhang, H.~Li, L.-X. You, Z.~Wang, J.~P. Dowling, S.~H\"ofling, C.-Y. Lu, and J.-W. Pan, ``Toward scalable boson sampling with photon loss,'' {\em Phys. Rev. Lett.}, vol.~120, p.~230502, Jun 2018.

\bibitem{flamini2018photonic}
F.~Flamini, N.~Spagnolo, and F.~Sciarrino, ``Photonic quantum information processing: a review,'' {\em arXiv:1803.02790}, 2018.

\bibitem{nevilleBristol2017}
A.~Neville, C.~Sparrow, R.~Clifford, E.~Johnston, P.~M. Birchall, A.~Montanaro, and A.~Laing, ``Classical boson sampling algorithms with superior performance to near-term experiments,'' {\em Nature Physics}, vol.~13, no.~12, p.~1153, 2017.

\bibitem{Clifford&Clifford}
P.~Clifford and R.~Clifford, ``The classical complexity of boson sampling,'' {\em arXiv:1706.01260}, 2017.

\bibitem{Clifford&Clifford_2020}
P.~Clifford and R.~Clifford, ``Faster classical boson sampling,'' {\em arXiv:2005.04214}, 2020.

\bibitem{kaneda2019}
F.~Kaneda and P.~G. Kwiat, ``High-efficiency single-photon generation via large-scale active time multiplexing,'' {\em Science advances}, vol.~5, no.~10, p.~eaaw8586, 2019.

\bibitem{loredo2016scalable}
J.~C. Loredo, N.~A. Zakaria, N.~Somaschi, C.~Anton, L.~De~Santis, V.~Giesz, T.~Grange, M.~A. Broome, O.~Gazzano, G.~Coppola, {\em et~al.}, ``Scalable performance in solid-state single-photon sources,'' {\em Optica}, vol.~3, no.~4, pp.~433--440, 2016.

\bibitem{lenzini2017active}
F.~Lenzini, B.~Haylock, J.~C. Loredo, R.~A. Abrahao, N.~A. Zakaria, S.~Kasture, I.~Sagnes, A.~Lemaitre, H.-P. Phan, D.~V. Dao, {\em et~al.}, ``Active demultiplexing of single photons from a solid-state source,'' {\em Laser \& Photonics Reviews}, vol.~11, no.~3, p.~1600297, 2017.

\bibitem{uppu2020scalable}
R.~Uppu, F.~T. Pedersen, Y.~Wang, C.~T. Olesen, C.~Papon, X.~Zhou, L.~Midolo, S.~Scholz, A.~D. Wieck, A.~Ludwig, and P.~L. Lodahl, ``Scalable integrated single-photon source,'' {\em Science advances}, vol.~6, no.~50, p.~eabc8268, 2020.

\bibitem{Patron_Renema}
R.~Garc{\'{i}}a-Patr{\'{o}}n, J.~J. Renema, and V.~Shchesnovich, ``Simulating boson sampling in lossy architectures,'' {\em {Quantum}}, vol.~3, p.~169, Aug. 2019.

\bibitem{Oszmaniec_Brod_err}
M.~Oszmaniec and D.~J. Brod, ``Classical simulation of photonic linear optics with lost particles,'' {\em arXiv:1801.06166}, 2018.

\bibitem{Renema_classical_simulability_arxiv}
J.~Renema, V.~Shchesnovich, and R.~Garcia-Patron, ``Classical simulability of noisy boson sampling,'' {\em arXiv:1809.01953}, 2018.

\bibitem{Lund2014}
A.~P. Lund, A.~Laing, S.~Rahimi-Keshari, T.~Rudolph, J.~L. O'Brien, and T.~C. Ralph, ``Boson sampling from a gaussian state,'' {\em Phys. Rev. Lett.}, vol.~113, p.~100502, Sep 2014.

\bibitem{gaussian_BS_prl_2017}
C.~S. Hamilton, R.~Kruse, L.~Sansoni, S.~Barkhofen, C.~Silberhorn, and I.~Jex, ``Gaussian boson sampling,'' {\em Phys. Rev. Lett.}, vol.~119, p.~170501, Oct 2017.

\bibitem{Chabaud2017}
U.~Chabaud, T.~Douce, D.~Markham, P.~van Loock, E.~Kashefi, and G.~Ferrini, ``Continuous-variable sampling from photon-added or photon-subtracted squeezed states,'' {\em Phys. Rev. A}, vol.~96, p.~062307, Dec 2017.

\bibitem{Lund2017measurement}
A.~P. Lund, S.~Rahimi-Keshari, and T.~C. Ralph, ``Exact boson sampling using gaussian continuous-variable measurements,'' {\em Phys. Rev. A}, vol.~96, p.~022301, Aug 2017.

\bibitem{Chakhmakhchyan2017}
L.~Chakhmakhchyan and N.~J. Cerf, ``Boson sampling with gaussian measurements,'' {\em Phys. Rev. A}, vol.~96, p.~032326, Sep 2017.

\bibitem{Gardiner1993}
C.~W. Gardiner, ``Driving a quantum system with the output field from another driven quantum system,'' {\em Phys. Rev. Lett.}, vol.~70, pp.~2269--2272, Apr 1993.

\bibitem{Carmichael1993}
H.~J. Carmichael, ``Quantum trajectory theory for cascaded open systems,'' {\em Phys. Rev. Lett.}, vol.~70, pp.~2273--2276, Apr 1993.

\bibitem{Aaronson2014}
S.~Aaronson and A.~Arkhipov, ``Bosonsampling is far from uniform,'' {\em Quantum Info. Comput.}, vol.~14, p.~1383–1423, Nov. 2014.

\bibitem{Shchesnovich2016}
V.~S. Shchesnovich, ``Universality of generalized bunching and efficient assessment of boson sampling,'' {\em Phys. Rev. Lett.}, vol.~116, p.~123601, Mar 2016.

\bibitem{Tichy2014}
M.~C. Tichy, K.~Mayer, A.~Buchleitner, and K.~M\o{}lmer, ``Stringent and efficient assessment of boson-sampling devices,'' {\em Phys. Rev. Lett.}, vol.~113, p.~020502, Jul 2014.

\bibitem{Note1}
The paper \cite {Chabaud2021Quantum} cites a previous version of the current paper previously available on arXiv.

\bibitem{Hangleiter2019}
D.~Hangleiter, M.~Kliesch, J.~Eisert, and C.~Gogolin, ``Sample complexity of device-independently certified ``quantum supremacy'','' {\em Phys. Rev. Lett.}, vol.~122, p.~210502, May 2019.

\bibitem{weedbrook2012gaussian}
C.~Weedbrook, S.~Pirandola, R.~Garc{\'\i}a-Patr{\'o}n, N.~J. Cerf, T.~C. Ralph, J.~H. Shapiro, and S.~Lloyd, ``Gaussian quantum information,'' {\em Reviews of Modern Physics}, vol.~84, no.~2, p.~621, 2012.

\bibitem{Furusawa_10000}
S.~Yokoyama, R.~Ukai, S.~C. Armstrong, C.~Sornphiphatphong, T.~Kaji, S.~Suzuki, J.-i. Yoshikawa, H.~Yonezawa, N.~C. Menicucci, and A.~Furusawa, ``Ultra-large-scale continuous-variable cluster states multiplexed in the time domain,'' {\em Nature Photonics}, vol.~7, no.~12, p.~982, 2013.

\bibitem{Furusawa_one_million}
J.-i. Yoshikawa, S.~Yokoyama, T.~Kaji, C.~Sornphiphatphong, Y.~Shiozawa, K.~Makino, and A.~Furusawa, ``Invited article: Generation of one-million-mode continuous-variable cluster state by unlimited time-domain multiplexing,'' {\em APL Photonics}, vol.~1, no.~6, p.~060801, 2016.

\bibitem{molecularBS_theory}
J.~Huh, G.~G. Guerreschi, B.~Peropadre, J.~R. McClean, and A.~Aspuru-Guzik, ``Boson sampling for molecular vibronic spectra,'' {\em Nature Photonics}, vol.~9, no.~9, p.~615, 2015.

\bibitem{molecularBS_exp}
C.~Sparrow, E.~Mart{\'\i}n-L{\'o}pez, N.~Maraviglia, A.~Neville, C.~Harrold, J.~Carolan, Y.~N. Joglekar, T.~Hashimoto, N.~Matsuda, J.~L. O’Brien, {\em et~al.}, ``Simulating the vibrational quantum dynamics of molecules using photonics,'' {\em Nature}, vol.~557, no.~7707, p.~660, 2018.

\bibitem{aspuru2012photonic}
A.~Aspuru-Guzik and P.~Walther, ``Photonic quantum simulators,'' {\em Nature Physics}, vol.~8, no.~4, p.~285, 2012.

\bibitem{georgescu2014quantum}
I.~Georgescu, S.~Ashhab, and F.~Nori, ``Quantum simulation,'' {\em Reviews of Modern Physics}, vol.~86, no.~1, p.~153, 2014.

\bibitem{verifying_NP}
J.~M. Arrazola, E.~Diamanti, and I.~Kerenidis, ``Quantum superiority for verifying np-complete problems with linear optics,'' {\em npj Quantum Information}, vol.~4, no.~1, p.~56, 2018.

\bibitem{MORDOR}
K.~R. Motes, J.~P. Olson, E.~J. Rabeaux, J.~P. Dowling, S.~J. Olson, and P.~P. Rohde, ``Linear optical quantum metrology with single photons: Exploiting spontaneously generated entanglement to beat the shot-noise limit,'' {\em Phys. Rev. Lett.}, vol.~114, p.~170802, Apr 2015.

\bibitem{BS_quantum_crypt}
Z.~Huang, P.~P. Rohde, D.~W. Berry, P.~Kok, J.~P. Dowling, and C.~Lupo, ``Photonic quantum data locking,'' {\em {Quantum}}, vol.~5, p.~447, Apr. 2021.

\bibitem{kok2010introduction}
P.~Kok and B.~W. Lovett, {\em Introduction to optical quantum information processing}.
\newblock Cambridge university press, 2010.

\bibitem{Roman_15db}
H.~Vahlbruch, M.~Mehmet, K.~Danzmann, and R.~Schnabel, ``Detection of 15 db squeezed states of light and their application for the absolute calibration of photoelectric quantum efficiency,'' {\em Phys. Rev. Lett.}, vol.~117, p.~110801, Sep 2016.

\bibitem{Menicicci2014}
N.~C. Menicucci, ``Fault-tolerant measurement-based quantum computing with continuous-variable cluster states,'' {\em Phys. Rev. Lett.}, vol.~112, p.~120504, Mar 2014.

\bibitem{GKP_encoding}
D.~Gottesman, A.~Kitaev, and J.~Preskill, ``Encoding a qubit in an oscillator,'' {\em Phys. Rev. A}, vol.~64, p.~012310, Jun 2001.

\bibitem{SPDC_general}
A.~Christ, A.~Fedrizzi, H.~H{\"u}bel, T.~Jennewein, and C.~Silberhorn, ``Parametric down-conversion,'' in {\em Experimental Methods in the Physical Sciences}, vol.~45, pp.~351--410, Elsevier, 2013.

\bibitem{Farella2023}
B.~Farella, G.~Medwig, R.~A. Abrahao, and A.~Nomerotski, ``{Spectral characterization of an SPDC source with a fast broadband spectrometer},'' {\em AIP Advances}, vol.~14, p.~045034, 04 2024.

\bibitem{PRL_Oxford_spdc2008}
P.~J. Mosley, J.~S. Lundeen, B.~J. Smith, P.~Wasylczyk, A.~B. U'Ren, C.~Silberhorn, and I.~A. Walmsley, ``Heralded generation of ultrafast single photons in pure quantum states,'' {\em Phys. Rev. Lett.}, vol.~100, p.~133601, Apr 2008.

\bibitem{Note2}
In the case of Scattershot Boson Sampling \cite {Lund2014}, $m$ refers to the number of two-mode squeezers, i.e., the number of SPDC single-photon sources. In that paper, the condition $m=n^2$ is imposed. In the present work, we use a more general definition: $m$ is simply the number of modes in the linear optical network represented by a $m$ x $m$ unitary matrix ($U$).

\bibitem{Grier2022}
D.~Grier, D.~J. Brod, J.~M. Arrazola, M.~B. d.~A. Alonso, and N.~Quesada, ``The {C}omplexity of {B}ipartite {G}aussian {B}oson {S}ampling,'' {\em {Quantum}}, vol.~6, p.~863, Nov. 2022.

\bibitem{gaussian_BS_arxiv_supp}
R.~Kruse, C.~S. Hamilton, L.~Sansoni, S.~Barkhofen, C.~Silberhorn, and I.~Jex, ``A detailed study of gaussian boson sampling,'' {\em arXiv:1801.07488}, 2018.

\bibitem{deshpande2022quantum}
A.~Deshpande, A.~Mehta, T.~Vincent, N.~Quesada, M.~Hinsche, M.~Ioannou, L.~Madsen, J.~Lavoie, H.~Qi, J.~Eisert, {\em et~al.}, ``Quantum computational advantage via high-dimensional gaussian boson sampling,'' {\em Science advances}, vol.~8, no.~1, p.~eabi7894, 2022.

\bibitem{BS_integrated_bristol_2019}
S.~Paesani, Y.~Ding, R.~Santagati, L.~Chakhmakhchyan, C.~Vigliar, K.~Rottwitt, L.~K. Oxenl{\o}we, J.~Wang, M.~G. Thompson, and A.~Laing, ``Generation and sampling of quantum states of light in a silicon chip,'' {\em Nature Physics}, pp.~1--5, 2019.

\bibitem{Time-Bin_BS_motes}
K.~R. Motes, A.~Gilchrist, J.~P. Dowling, and P.~P. Rohde, ``Scalable boson sampling with time-bin encoding using a loop-based architecture,'' {\em Phys. Rev. Lett.}, vol.~113, p.~120501, Sep 2014.

\bibitem{leonhardt1997measuring}
U.~Leonhardt, {\em Measuring the quantum state of light}.
\newblock Cambridge university press, 1997.

\bibitem{Bachor&Ralph}
H.-A. Bachor and T.~C. Ralph, {\em A guide to experiments in quantum optics}.
\newblock Wiley, 2004.

\bibitem{Aolita2015}
L.~Aolita, C.~Gogolin, M.~Kliesch, and J.~Eisert, ``Reliable quantum certification of photonic state preparations,'' {\em Nature Communications}, vol.~6, nov 2015.

\bibitem{Chabaud2022}
U.~Chabaud, A.~Deshpande, and S.~Mehraban, ``Quantum-inspired permanent identities,'' {\em {Quantum}}, vol.~6, p.~877, Dec. 2022.

\bibitem{webb2006homodyne}
J.~G. Webb, T.~C. Ralph, and E.~H. Huntington, ``Homodyne measurement of the average photon number,'' {\em Physical Review A}, vol.~73, no.~3, p.~033808, 2006.

\bibitem{PhysRevApplied2021_Asavanant}
W.~Asavanant, B.~Charoensombutamon, S.~Yokoyama, T.~Ebihara, T.~Nakamura, R.~N. Alexander, M.~Endo, J.-i. Yoshikawa, N.~C. Menicucci, H.~Yonezawa, and A.~Furusawa, ``Time-domain-multiplexed measurement-based quantum operations with 25-mhz clock frequency,'' {\em Phys. Rev. Appl.}, vol.~16, p.~034005, Sep 2021.

\bibitem{larsen2021deterministic}
M.~V. Larsen, X.~Guo, C.~R. Breum, J.~S. Neergaard-Nielsen, and U.~L. Andersen, ``Deterministic multi-mode gates on a scalable photonic quantum computing platform,'' {\em Nature Physics}, vol.~17, no.~9, pp.~1018--1023, 2021.

\bibitem{larsen2019Science}
M.~V. Larsen, X.~Guo, C.~R. Breum, J.~S. Neergaard-Nielsen, and U.~L. Andersen, ``Deterministic generation of a two-dimensional cluster state,'' {\em Science}, vol.~366, no.~6463, pp.~369--372, 2019.

\bibitem{asavanant2019Science}
W.~Asavanant, Y.~Shiozawa, S.~Yokoyama, B.~Charoensombutamon, H.~Emura, R.~N. Alexander, S.~Takeda, J.-i. Yoshikawa, N.~C. Menicucci, H.~Yonezawa, {\em et~al.}, ``Generation of time-domain-multiplexed two-dimensional cluster state,'' {\em Science}, vol.~366, no.~6463, pp.~373--376, 2019.

\bibitem{wilde_2013}
M.~M. Wilde, {\em Quantum Information Theory}.
\newblock Cambridge University Press, 2013.

\bibitem{nielsen_chuang_book}
M.~A. Nielsen and I.~L. Chuang, {\em Quantum computation and Quantum Information}.
\newblock Cambridge University Press, 2000.

\bibitem{fuchs1997cryptographic}
C.~Fuchs and J.~van~de Graaf, ``Cryptographic distinguishability measures for quantum-mechanical states,'' {\em IEEE Transactions on Information Theory}, vol.~45, no.~4, pp.~1216--1227, 1999.

\bibitem{zhang2018IEEE}
X.~Zhang, Y.~Zhang, Z.~Li, S.~Yu, and H.~Guo, ``1.2-ghz balanced homodyne detector for continuous-variable quantum information technology,'' {\em IEEE Photonics Journal}, vol.~10, no.~5, pp.~1--10, 2018.

\bibitem{Lita_NIST_OE_TES}
A.~E. Lita, A.~J. Miller, and S.~W. Nam, ``Counting near-infrared single-photons with 95\% efficiency,'' {\em Opt. Express}, vol.~16, pp.~3032--3040, Mar 2008.

\bibitem{TES_UQ_arxiv}
L.~A. Morais, T.~Weinhold, M.~P. de~Almeida, J.~Combes, A.~Lita, T.~Gerrits, S.~W. Nam, A.~G. White, and G.~Gillett, ``Precisely determining photon-number in real-time,'' {\em arXiv:2012.10158}, 2020.

\bibitem{NatPht_2022_TES}
M.~Eaton, A.~Hossameldin, R.~J. Birrittella, P.~M. Alsing, C.~C. Gerry, H.~Dong, C.~Cuevas, and O.~Pfister, ``Resolution of 100 photons and quantum generation of unbiased random numbers,'' {\em Nature Photonics}, vol.~17, 2023.

\bibitem{PRL2019_Lewis}
L.~A. Howard, G.~G. Gillett, M.~E. Pearce, R.~A. Abrahao, T.~J. Weinhold, P.~Kok, and A.~G. White, ``Optimal imaging of remote bodies using quantum detectors,'' {\em Phys. Rev. Lett.}, vol.~123, p.~143604, Sep 2019.

\bibitem{gerrits_tes_intro}
T.~Gerrits, A.~Lita, B.~Calkins, and S.~W. Nam, ``Superconducting transition edge sensors for quantum optics,'' in {\em Superconducting devices in quantum optics}, pp.~31--60, Springer, 2016.

\bibitem{Ottawa_loop}
N.~M. Sullivan, B.~Braverman, J.~Upham, and R.~W. Boyd, ``Photon number resolving detection with a single-photon detector and adaptive storage loop,'' {\em arXiv:2311.13515}, 2023.

\bibitem{nanowire_number_resolving}
C.~Cahall, K.~L. Nicolich, N.~T. Islam, G.~P. Lafyatis, A.~J. Miller, D.~J. Gauthier, and J.~Kim, ``Multi-photon detection using a conventional superconducting nanowire single-photon detector,'' {\em Optica}, vol.~4, pp.~1534--1535, Dec 2017.

\bibitem{PhysRevApplied_number_resolving}
K.~L. Nicolich, C.~Cahall, N.~T. Islam, G.~P. Lafyatis, J.~Kim, A.~J. Miller, and D.~J. Gauthier, ``Universal model for the turn-on dynamics of superconducting nanowire single-photon detectors,'' {\em Phys. Rev. Appl.}, vol.~12, p.~034020, Sep 2019.

\bibitem{Aaronson_Brod_error}
S.~Aaronson and D.~J. Brod, ``Bosonsampling with lost photons,'' {\em Physical Review A}, vol.~93, no.~1, p.~012335, 2016.

\bibitem{brod_oszmaniec_nonuniform_losses}
D.~J. Brod and M.~Oszmaniec, ``Classical simulation of linear optics subject to nonuniform losses,'' {\em {Quantum}}, vol.~4, p.~267, May 2020.

\bibitem{Arkhipov_error_analysis}
A.~Arkhipov, ``Bosonsampling is robust against small errors in the network matrix,'' {\em Phys. Rev. A}, vol.~92, p.~062326, Dec 2015.

\bibitem{rohde_ralph_error}
P.~P. Rohde and T.~C. Ralph, ``Error tolerance of the boson-sampling model for linear optics quantum computing,'' {\em Physical Review A}, vol.~85, no.~2, p.~022332, 2012.

\bibitem{L_Patron_error}
A.~Leverrier and R.~Garc{\'\i}a-Patr{\'o}n, ``Analysis of circuit imperfections in bosonsampling,'' {\em Quantum Information \& Computation}, vol.~15, pp.~489--512, 2015.

\bibitem{rahimi_ralph_caves_sufficient}
S.~Rahimi-Keshari, T.~C. Ralph, and C.~M. Caves, ``Sufficient conditions for efficient classical simulation of quantum optics,'' {\em Physical Review X}, vol.~6, no.~2, p.~021039, 2016.

\bibitem{takeda2017universal}
S.~Takeda and A.~Furusawa, ``Universal quantum computing with measurement-induced continuous-variable gate sequence in a loop-based architecture,'' {\em Physical review letters}, vol.~119, no.~12, p.~120504, 2017.

\bibitem{spie_travelling_wave_of_light}
T.~Serikawa, Y.~Shiozawa, H.~Ogawa, N.~Takanashi, S.~Takeda, J.-i. Yoshikawa, and A.~Furusawa, ``Quantum information processing with a travelling wave of light,'' in {\em Integrated Optics: Devices, Materials, and Technologies XXII}, vol.~10535, p.~105351B, International Society for Optics and Photonics, 2018.

\bibitem{furusawa_review_apl_2019}
S.~Takeda and A.~Furusawa, ``Toward large-scale fault-tolerant universal photonic quantum computing,'' {\em APL Photonics}, vol.~4, no.~6, p.~060902, 2019.

\end{thebibliography}

\clearpage
\widetext
\section*{Appendix}

\subsection*{Diagonal Covariance Matrices}
  The diagonal covariance matrix $\Sigma$ for the simulation in Fig. \ref{Chernoff_bound_plots} for $m = 4$ is:
  \begin{equation*}
      \Sigma = \text{diag}(10,0.1, \cdots, 10, 0.1)_{16\times 16} = \begin{bmatrix}
      10 & 0 & \cdots & 0 & 0 \\
      0 & 0.1 & \cdots & 0 & 0 \\
      \vdots & \vdots &\ddots & \vdots & \vdots \\
      0 & 0 & \cdots  & 10 & 0 \\
      0 & 0 & \cdots & 0 & 0.1
      \end{bmatrix}_{16\times 16}
  \end{equation*}
  The diagonal covariance matrix $\Sigma$ for the simulation in Fig. \ref{Chernoff_bound_plots} for $m = 9$ is:
   \begin{equation*}
      \Sigma = \text{diag}(10,0.1, \cdots, 10, 0.1)_{36\times 36} = \begin{bmatrix}
      10 & 0 & \cdots & 0 & 0 \\
      0 & 0.1 & \cdots & 0 & 0 \\
      \vdots & \vdots &\ddots & \vdots & \vdots \\
      0 & 0 & \cdots  & 10 & 0 \\
      0 & 0 & \cdots & 0 & 0.1
      \end{bmatrix}_{36\times 36}
  \end{equation*}
\subsection*{Output State Characterization Scaling: Numerical Simulations}
	
A Gaussian state can be fully characterized by the mean vector (which we will assume zero) and the covariance matrix. Therefore, it suffices to estimate the $ 4m \times 4m $ covariance matrix of the Gaussian state prior to single-photon detection. 
	
This can be done using CV homodyne detection. To estimate the scaling with the number of dual-homodyne measurement samples needed for the sample average to be "close" to the true average, one can apply the operator Chernoff bound. It provides a lower bound for the probability that the sample average does not deviate significantly from the true average. 
	
In the main text, we present three versions of the operator Chernoff bound. 
\begin{gather}
		\text{Pr}\Big\{ \frac{(1-\eta)}{2C^2} (\Sigma + I) \leq \bar{\xi} \leq \frac{(1+\eta)}{2C^2} (\Sigma + I) \Big\} \nonumber \\
		\geq 1 - 8 m e^{-K \eta^2 /(8C^2\ln 2)} \label{eq:12}\\
		\text{Pr}\Big\{\frac{(1-\eta)\Sigma - \eta I}{C^2} \leq 2\bar{\xi} - \frac{I}{C^2} \leq \frac{(1+\eta)\Sigma + \eta I}{C^2}\Big\}  \nonumber\\
		\geq 1 - 8 m e^{-K \eta^2 /(8C^2\ln 2)} \label{eq:13}\\
		\text{Pr}\Big\{\frac{[1-\eta(1+b^{-1})]\Sigma}{C^2} \leq 2\bar{\xi} - \frac{I}{C^2} \leq \frac{[1+\eta(1+b^{-1})]\Sigma}{C^2}\Big\} \nonumber\\
		\geq 1 - 8 m e^{-K \eta^2 /(8C^2\ln 2)}  \label{eq:14}
	\end{gather}
    In Eq. \ref{eq:12}, the operator Chernoff bound is written for $\frac{1}{2C^2}(\Sigma + I)$ (includes half a unit of vacuum noise). Equations \ref{eq:13} and \ref{eq:14} describe the operator Chernoff bound for the scaled covariance matrix $ \Sigma/C^2 $. 
	
    One can numerically demonstrate the validity of these inequalities. To remind ourselves, $ K $ is the number of samples, $ m $ is the number of modes, $b$ is the variance of the quadrature for the maximum possible squeezing, and $ C $ is a scaling factor for the CV vector samples. The steps used in the numerical simulation are as follows.
	\begin{enumerate}
		\item We draw $ K $ $ 4m $-dimensional vector samples from a multivariate Gaussian distribution with mean-vector zero and covariance matrix $ (\Sigma + I)/2 $.
		\begin{equation*}
			s_i \sim \mathcal{N}_{4m} \left(\mathbf{0},  (\Sigma + I)/2\right) 
		\end{equation*}
  \item We calculate the matrix samples from the outer product of each scaled vector sample, $s_i/C$, with itself.
  \begin{equation*}
      \xi_i = \left(\frac{s_i}{C}\right)\left(\frac{s^T_i}{C}\right) = \frac{1}{C^2} s_i s^T_i
  \end{equation*}
    \item We check that each $\xi_i$ has all of its eigenvalues in $[0,1]$. This condition is necessary for applying the operator Chernoff bound. If this fails, we restart from step 1. A larger scaling factor $C$ gives a higher success rate. However, a smaller $C$ value provides a tighter bound.
		\item We calculate the sample average
		\begin{equation*}
			\bar{\xi} = \frac{1}{K} \sum_{i=1}^{K} \xi_i 
		\end{equation*} 
		\item We perform multiple runs of steps 1-4 (typically 50 runs). We then obtain a list $ \bar{z} $ of all the sample averages:
		\begin{equation*}
			\bar{z} = [\bar{\xi}_1, \bar{\xi}_2, \cdots, \bar{\xi}_{\text{runs}}]
		\end{equation*}
		\item For each $ \eta \in (0,0.5) $ (step size $\Delta \eta = 0.001 $), we use each $ \bar{\xi}_j \in \bar{z} $ to see if the two conditions of the operator Chernoff bound under consideration are true. For example, for Eq. \ref{eq:14}, we would check that
		\begin{gather}
			\frac{[1-\eta(1+b^{-1})]\Sigma}{C^2} \leq 2\bar{\xi} - \frac{I}{C^2} \nonumber\\
			\iff 2\bar{\xi} - \frac{I}{C^2} -\frac{[1-\eta(1+b^{-1})]\Sigma}{C^2} \geq 0 \label{eq:15}
		\end{gather}
		and 
		\begin{gather}
			2\bar{\xi} - \frac{I}{C^2} \leq \frac{[1+\eta(1+b^{-1})]\Sigma}{C^2} \nonumber\\
			\iff \frac{[1+\eta(1+b^{-1})]\Sigma}{C^2} - 2\bar{\xi} + \frac{I}{C^2} \geq 0 \label{eq:16}.
		\end{gather}
	This can be done by verifying whether all of the eigenvalues of the matrix on the left-hand side of Eq. \ref{eq:15}, and all of the eigenvalues of the matrix on the left-hand side of Eq. \ref{eq:16} are non-negative. As such, we obtain a corresponding boolean list for $ \bar{z} $. 
	\item The probability on the left-hand side of the operator Chernoff bound under consideration is calculated for each $ \eta $ as the number of true cases divided by the total number of runs. For instance, for Eq. \ref{eq:14}:
\begin{equation*}
    \text{Pr}\Big\{\frac{[1-\eta(1+b^{-1})]\Sigma}{C^2} \leq 2\bar{\xi} - \frac{I}{C^2} \leq \frac{[1+\eta(1+b^{-1})]\Sigma}{C^2}\Big\} = \frac{\text{\# of True cases} (\bar{z};\eta)}{\text{runs}}
\end{equation*}
	\item The right-hand side of the operator Chernoff bound under consideration is calculated (using the analytic expression) for all $ \eta \in (0, 0.5) $ as
	\begin{equation*}
		1 - 8 m e^{-K\eta^2/(8C^2 \ln 2)}
	\end{equation*}
	\item Both sides of the operator Chernoff bound under consideration are plotted as a function of $ \eta $.  
 \end{enumerate}
Note that it is important for the scaling factor $ C $ to be chosen large enough such that each $ \xi_i $ has all of its eigenvalues between zero and one:
	\begin{equation*}
		\forall i \in [K]: 0 \leq \xi_i \leq I.
	\end{equation*}
	This is a necessary condition for applying the operator Chernoff bound. One can check if the chosen scaling parameter $ C $ ensures that the condition above holds with high enough probability. This can be verified by creating a histogram of the top eigenvalues of $ \xi_i$. Although a smaller scaling $C$ (with other parameters being fixed) can provide a tighter bound, the required condition above for applying the bound would be satisfied at an increasingly infrequent rate. This can be estimated as the joint probability for all of the $K$ samples $\xi_i$ to have their eigenvalues in $[0,1]$.  
    \subsection*{Non-diagonal Covariance Matrix Simulations}
    In addition to the results in the main paper, we also performed this simulation for non-diagonal covariance matrices of the form
    \begin{equation*}
        \Sigma = O^T \Sigma_\text{diag} O,
    \end{equation*}
    where $O$ is an arbitrary orthogonal matrix and $\Sigma_\text{diag}$ is the diagonal covariance matrix originally chosen. This preserves the eigenvalues of the original covariance matrix $\Sigma_\text{diag}$. 
 
\begin{figure}[H]
    \centering
    \includegraphics[width =8cm]{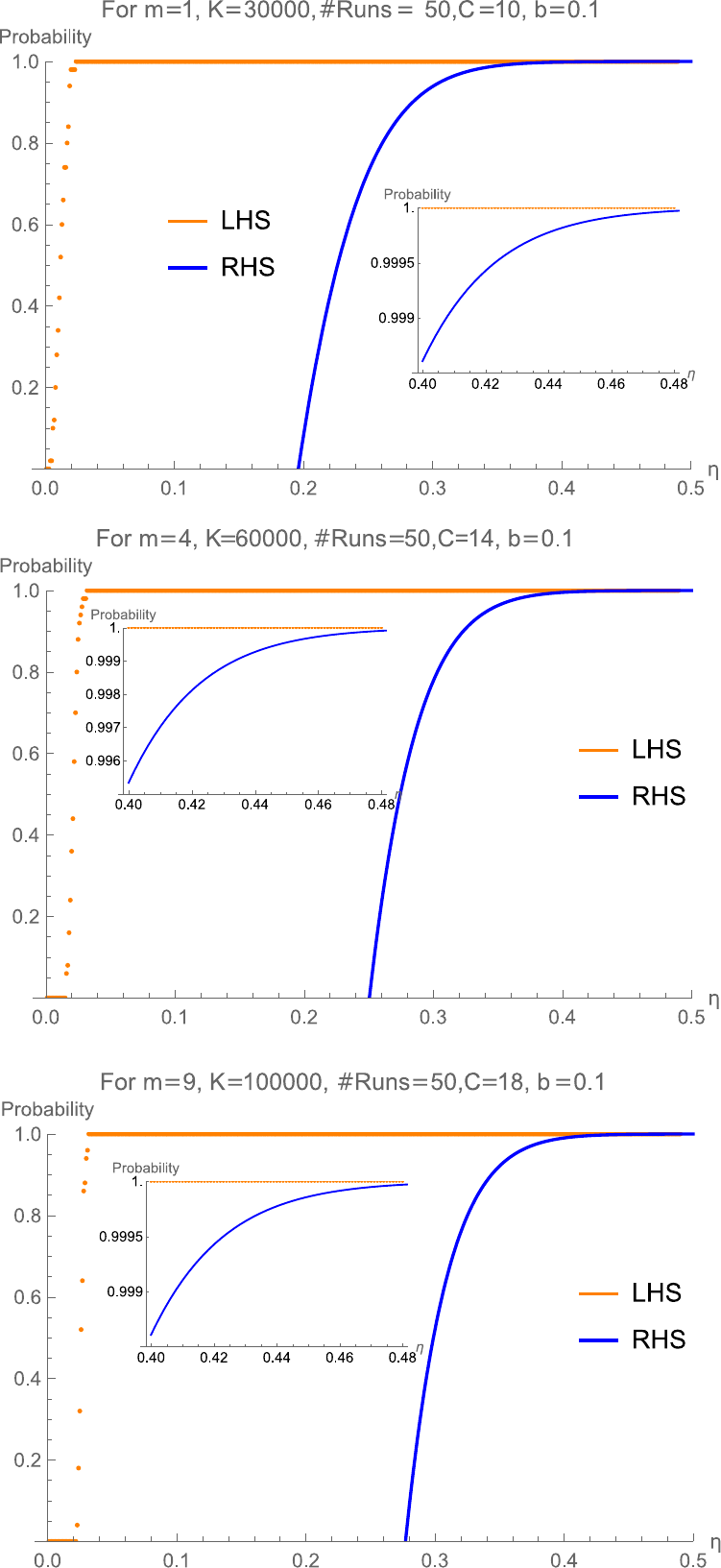}
    \caption{Plots of the probabilities on the left-hand side (LHS, shown as orange dots) and the right-hand side (RHS, shown as solid blue) of Eq. \ref{eq:14} as a function of $\eta$ for the specified parameters. These are for the non-diagonal matrices specified. The insets show the region $\eta \geq 0.4$. The simulations are in agreement with Eq. \ref{Chernoff_eq}.}
    \label{Chernoff_bound_plots_supp2}
\end{figure}

    In Fig. \ref{Chernoff_bound_plots_supp2}, we present plots of the probabilities in Eq. \ref{eq:14} for examples of non-diagonal covariance matrices (see next page for the matrices). As in the diagonal case, we see agreement with the operator Chernoff bound Eq. \ref{eq:14}.  
\newpage
\subsection*{Non-diagonal Covariance Matrices}
The covariance matrix $\Sigma$ for the simulation in Fig. \ref{Chernoff_bound_plots_supp2} for $m = 1$ is:
\begin{equation*}
    \Sigma=\left.\left[\begin{array}{rrrr}3.16687&3.36663&-2.60207&-1.6882\\3.36683&7.67233&1.96645&-1.55674\\-2.60207&1.96645&8.30882&1.80136\\-1.6882&-1.55674&1.80136&1.05198\end{array}\right.\right]
\end{equation*}
The covariance matrix $\Sigma$ for the simulation in Fig. \ref{Chernoff_bound_plots_supp2} for $m = 4$ is:
\begin{figure}[H]
\centering
\rotatebox{-90}{{\includegraphics[scale=1,trim={1.5cm 0 0 0.5cm},clip]{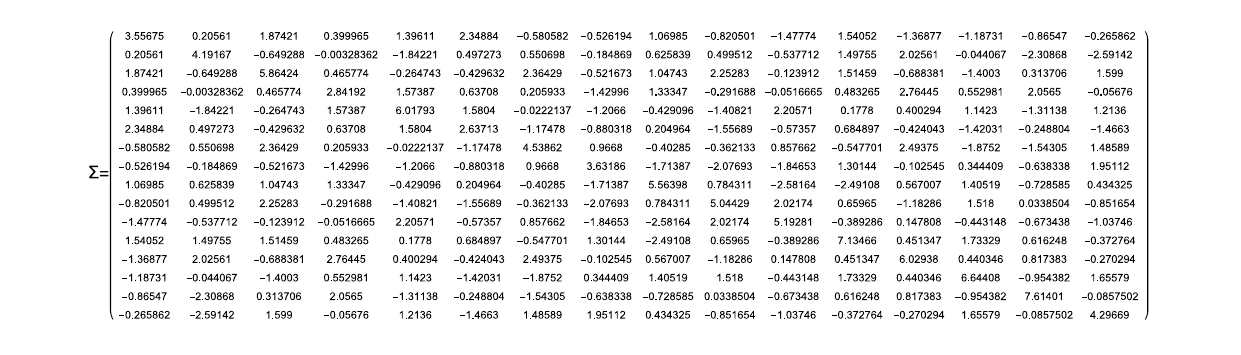}}}
\end{figure}
\newpage
The non-diagonal covariance matrix $\Sigma$ for the simulation in Fig. \ref{Chernoff_bound_plots_supp2} for $m = 9$ is:
\begin{figure}[H]
\centering
\rotatebox{-90}{{\includegraphics[scale=2.1,trim={1cm 0 0.5cm 0},clip]{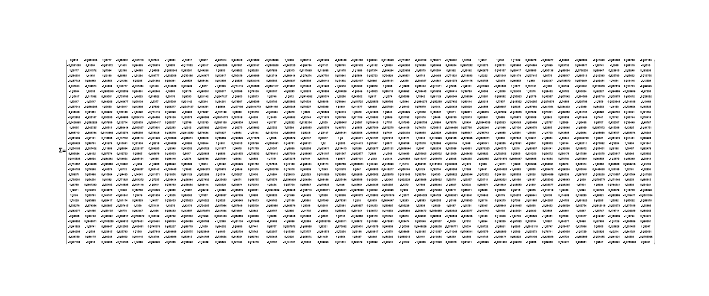}}}
\end{figure}
\subsection*{Numerical Factors for the Certification Scaling}

According to the all-vacuum certification process of Aolita \emph{et al.} \cite{Aolita2015}, for a Gaussian state with assumed zero mean, the number of samples $N$ needed for certification is required to satisfy
\begin{equation}
    N \geq 2^5 \frac{\sigma_2^2 (2 \kappa m + 1) m^2 s_{max}^4 \kappa}{\varepsilon^2 \ln\left(\frac{1}{1-\alpha}\right)},
\end{equation}
where $\varepsilon^2$ is an acceptable error range for the fidelity and $\alpha$ is the maximum probability that the fidelity lies outside this range. The other parameters relate to the state and are generally viewed through the lens of the Euler decomposition of the symplectic matrix $S$ used to generate the state. Also, $m$ is the number of modes of the state, and $\kappa$ is either $2 m$, or $2 d^2$ where $d$ is the maximum number of output modes that any input mode is connected to in the orthogonal matrix of the Euler decomposition. $\kappa$ is the smaller of these two options.  $s_{max}$ is the maximum value of the diagonal matrix in the Euler decomposition.  $\sigma_2$ is not particularly concerned with the Euler decomposition \textit{per se}, but relates to the second moments of the quadratures.  One must estimate these quantities from the actual data, and $\sigma_2$ is an upper bound on the variance of the second moment estimate.

The hardest bits of this equation are $\sigma_2$ and $s_{max}$.  Considering the scattershot sampling case, it is the case that the maximum probability of $n$ photons in $m$ mode pairs is given by a squeezing
\begin{equation}
    \tanh{r} = \frac{\sqrt{n}}{\sqrt{n+m}}
\end{equation}

with a probability that scales as $\frac{1}{\sqrt{n}}$ for $m=n^2$.  This will give us the expression for $s_{max}$ as the squeezing is exactly that used in the Euler decomposition.  So we can use $s_{max} = e^r$.  This asymptotes to $1$ and can therefore be easily upper bounded by some chosen point for low values of $n$.

Now turning to $\sigma_2$.  Our estimation of the covariance matrix elements is performed by the outer product of the sample vector.  A single replicate in the estimator is formed from the product of two components of the measured variables, say $r_i$ and $r_j$.  The quantity we are interested in is
\newcommand{\var}{\ensuremath\textrm{var}}
\newcommand{\cov}{\ensuremath\textrm{cov}}
\begin{equation}
\var(r_i r_j) = \cov(r_i^2,r_j^2) + 
\langle r_i^2 \rangle \langle r_j^2 \rangle - \cov(r_i, r_j)^2,
\end{equation}
where the RHS is assuming zero means. The last term, $\cov(r_i, r_j)$ is read off from the covariance matrix.  The first two terms, $\cov(r_i^2,r_j^2) + \langle r_i^2 \rangle \langle r_j^2 \rangle = \langle r_i^2 r_j^2\rangle$ are a higher order moment that we can reduce, for Gaussian states, using the Wick rules to $\langle r_i^2 \rangle \langle r_j^2 \rangle + 2 \langle r_i r_j \rangle^2 = \var(r_i)\var(r_j) + 2 \cov(r_i,r_j)^2$.  Putting it all together gives
\begin{equation}
\var(r_i r_j) = \var(r_i)\var(r_j) + \cov(r_i,r_j)^2.
\end{equation}
We are essentially interested in the maximum value for this second moment variance, and so we consider maximum values for these quantities.  The covariances are upper bounded by the variances, and the variances are upper bounded by the squeezing in the state.  The maximum variance in the state will be given by $(\cosh{2r})/2$ and hence we find that for the single replicate we have
\begin{equation}
\var(r_i r_j) \leq \frac{(\cosh{2r})^2}{2}.
\end{equation}
Considering the scaling in $n$, this would be bounded by a constant (greater than $2$), as $r$ decreases with increasing $n$.

For the certification of the squeezed vacuum states used as input for Boson Sampling, we have $\kappa = 2m$,\\ $r = \arctanh{\sqrt{\frac{\sqrt{m}}{(m+\sqrt{m})}}}$, and $\sigma_2 = \cosh^2(2r)/2$. Additionally, $s_\text{max} = e^r$, and we take $\alpha = \epsilon = 0.1$. We find the following condition for the required number of samples $N$:
\begin{equation}
    N \geq e^{4r} \cosh^4(2r)[(6.07438 \times 10^4) m^5 + (1.5186 \times 10^4)m^3].
    \label{eq:23}
\end{equation}
For $m \geq 25$, we have that
\begin{equation}
     e^{4r} \cosh^4(2r) \leq 21.76
     \label{eq:24}
\end{equation}
We can retrieve the simplified condition on the required number of samples in Eq. \ref{eq:certification} for $m\geq 25$ by combining Eq. \ref{eq:23} and Eq. \ref{eq:24}:
\begin{equation}
    N \geq (1.32158 \times 10^6) \ m^5 + (3.30395 \times 10^5) \ m^3
\end{equation}

\end{document}